\documentstyle[epsfig,graphics]{aipproc}
\title{Photon collider at TESLA: parameters  and interaction region issues }
\author{Valery Telnov~\thanks{e-mail:telnov@inp.nsk.su, telnov@mail.desy.de}}
\address{ Institute of Nuclear Physics, 630090, Novosibirsk, Russia \\
      and DESY, Germany \\ }
\begin{document}
\newcommand{\EP}{\mbox{e$^+$}}
\newcommand{\EM}{\mbox{e$^-$}}
\newcommand{\EPEM}{\mbox{e$^+$e$^-$}}
\newcommand{\EMEM}{\mbox{e$^-$e$^-$}}
\newcommand{\EE}{\mbox{ee}}
\newcommand{\GG}{\mbox{$\gamma\gamma$}}
\newcommand{\GP}{\mbox{$\gamma$e$^+$}}
\newcommand{\GE}{\mbox{$\gamma$e}}
\newcommand{\LGE}{\mbox{$L_{\GE}$}}
\newcommand{\LGG}{\mbox{$L_{\GG}$}}
\newcommand{\LEE}{\mbox{$L_{\EE}$}}
\newcommand{\TEV}{\mbox{TeV}}
\newcommand{\WGG}{\mbox{$W_{\gamma\gamma}$}}
\newcommand{\GEV}{\mbox{GeV}}
\newcommand{\EV}{\mbox{eV}}
\newcommand{\CM}{\mbox{cm}}
\newcommand{\M}{\mbox{m}}
\newcommand{\MM}{\mbox{mm}}
\newcommand{\NM}{\mbox{nm}}
\newcommand{\MKM}{\mbox{$\mu$m}}
\newcommand{\E}{\mbox{$\epsilon$}}
\newcommand{\EN}{\mbox{$\epsilon_n$}}
\newcommand{\EI}{\mbox{$\epsilon_i$}}
\newcommand{\ENI}{\mbox{$\epsilon_{ni}$}}
\newcommand{\ENX}{\mbox{$\epsilon_{nx}$}}
\newcommand{\ENY}{\mbox{$\epsilon_{ny}$}}
\newcommand{\EX}{\mbox{$\epsilon_x$}}
\newcommand{\EY}{\mbox{$\epsilon_y$}}
\newcommand{\SEC}{\mbox{s}}
\newcommand{\CMS}{\mbox{cm$^{-2}$s$^{-1}$}}
\newcommand{\MRAD}{\mbox{mrad}}
\newcommand{\IND}{\hspace*{\parindent}}
\newcommand{\beq}{\begin{equation}}
\newcommand{\eeq}{\end{equation}}
\newcommand{\beqn}{\begin{eqnarray}}
\newcommand{\eeqn}{\end{eqnarray}}
\newcommand{\dst}{\displaystyle}
\newcommand{\bm}{\boldmath}
\newcommand{\BX}{\mbox{$\beta_x$}}
\newcommand{\BY}{\mbox{$\beta_y$}}
\newcommand{\BI}{\mbox{$\beta_i$}}
\newcommand{\SX}{\mbox{$\sigma_x$}}
\newcommand{\SY}{\mbox{$\sigma_y$}}
\newcommand{\SZ}{\mbox{$\sigma_z$}}
\newcommand{\SI}{\mbox{$\sigma_i$}}
\newcommand{\SIP}{\mbox{$\sigma_i^{\prime}$}}
\newcommand{\n}{\mbox{$n_f$}}
\def\ZP #1 #2 #3 {{\it Z.\ Phys.}\ {\bf #1}\ (#2) #3}
\def\PR #1 #2 #3 {{\it Phys.\ Rev.}\ {\bf #1}\ (#2) #3}
\def\b{\beta}
\def\g{\gamma}
\def\SM{$\mathcal{SM}$}
\def\MSSM{$\mathcal{MSSM}$}
\def\2HDM{$2\mathcal{HDM}$}
\def\gg{\gamma\gamma}
\def\h{\rm h}
\def\ccbar{\overline{\mbox c}\mbox{c}}
\def\bbbar{\overline{\mbox b}\mbox{b}}
\def\qqbar{\overline{\mbox q}\mbox{q}}
\def\ccbarg{\overline{\mbox c}\mbox{cg}}
\def\bbbarg{\overline{\mbox b}\mbox{bg}}
\def\BR{\rm BR}
\newcommand{\sw}{\mbox{$\sin\Theta_W\,$}}
\newcommand{\cw}{\mbox{$\cos\Theta_W\,$}}
\newcommand{\epe}{\mbox{$e^+e^-\,$}}
\newcommand{\ggam}{\mbox{$\gamma\gamma\,$}}
\newcommand{\egam}{\mbox{$e\gamma\,$}}
\newcommand{\gewnu}{\mbox{$e\gamma\to W\nu\,$}}
\newcommand{\eeww}{\mbox{$e^+e^-\to W^+W^-\,$}}
\newcommand{\ggww}{\mbox{$\gamma\gamma\to W^+W^-\,$}}
\newcommand{\ggzz}{\mbox{$\gamma\gamma\to ZZ\,$}}
\newcommand{\egeh}{\mbox{$e\gamma\to eH\,$}}
\newcommand{\geeww}{\mbox{$e\gamma\to e W^+W^-\,$}}
\maketitle
\begin{abstract}
 
  Photon colliders (\GG, \GE) are based on backward Compton scattering
  of laser light off the high energy electrons of linear colliders.
  Recent study has shown that the \GG\ luminosity in the high energy
  peak can reach 0.3--0.5 $L_{\EPEM\ }$.  Typical cross sections of
  interesting processes in \GG\ collisions are higher than those in
  \EPEM collisions by about one order of magnitude, so the number of
  events in \GG\ collisions will be more than that in \EPEM\ 
  collisions.  In this paper  possible parameters of a photon
  collider at TESLA and a laser scheme are briefly discussed.

\end{abstract}
\section{Introduction}

The unique feature of the $e^+e^-$ Linear Colliders (LC) with the
energy from hundreds GeV to several TeV is the possibility to
construct on its basis a Photon Linear Collider (PLC) using the
process of the Compton backscattering of laser light off the high
energy electrons~\cite{GKST83}.

The maximum c.m.s.  energy in \GG\ collisions reaches about 0.8 (0.9
in \GE\ collisions) of that in \EPEM\ collisions.  Typical luminosity
distribution~\cite{Tfrei,GGTEL} in \GG\ collisions has a high energy peak
and some low energy part.  The peak has the width at half of maximum
about 15\%, photons here can have high degree of circular
polarization.  This region is the most valuable for experimentation.
Comparing event rates in \GG\ and \EPEM\ collisions we will use the
value of \GG\ luminosity in this peak.

In this talk I briefly discuss physics, possible parameters of the
photon collider at TESLA and a lasers--optics scheme.
For more details and references see my recent paper~\cite{GGTEL}.

\section{Physics}

 Physics in \EPEM\ and \GG, \GE\ collisions is quite similar
because the same particles can be produced. However, reactions are
different and can give complementary information. Some phenomena can
best be studied at photon colliders due to better accuracy (larger
cross-sections) or larger accessible masses (a single resonance (in
\GG\ and \GE) or a pair of light and heavy particles (in \GE).
A short list of processes for the physics program of the photon collider 
is presented in Table \ref{processes} ~\cite{summary}.
\begin{table}[!hbtp]
\caption{Gold-plated processes at photon colliders}
\vspace{0mm}
{\renewcommand{\arraystretch}{1.2}
\begin{center}
\begin{tabular}{| l | c |} 
$\g\g\to h^0 \to b\bar b$ & \SM\ Higgs, $m_{h^0}<160$~GeV or \MSSM\ Higgs, $m_{h^0}<130$~GeV\\
$\g\g\to h^0 \to WW^*$    & \SM\ Higgs,
140~GeV$<m_{h^0}<190$~GeV \\
$\g\g\to h^0 \to ZZ$      & \SM\
Higgs, 180~GeV$<m_{h^0}<350$~GeV \\
$\g\g\to H,A\to b\bar b$  &
single production of \MSSM\ heavy Higgs states, for large
$\tan\beta$\\
$\g\g\to \tilde{f}\bar{\tilde{f}},\
\tilde{\chi}^+_i\tilde{\chi}^-_i,\ H^+H^-$& large cross sections,
possible observations of FCNC
\\ $\g\g\to S$ &
$\tilde{t}\bar{\tilde{t}}$ stoponium in the $S$-wave \\
$e^-\gamma \to \tilde{e}^- \tilde{\chi}_1^0$ & $m_{\tilde{e}^-} < 0.9 \sqrt{s_{e^+e^-}} - m_{\tilde{\chi}_1^0}$\\
$\g\g\to W^+W^-$ & anomalous $W$ interactions, extra dimensions \\
$\g e^-\to W^- \nu_e$ & anomalous $W$ couplings \\
$\ggam\to WWWW$,$WWZZ$& strong $WW$ scattering, 
quartic anomalous $W$, $Z$  couplings\\
$\g\g\to t\bar t$ & anomalous top quark interactions \\
$\g e^-\to \bar t b \nu_e$ & anomalous $Wtb$ coupling \\
$e^-\g \to e^- X$ & 
     spin independent and spin dependent photon structure functions\\ 
$e^-\g \to \nu_e X$ & flavour decomposition of the quark distributions  
     in the photon\\
$\g g\to q\bar q,\ c\bar c$ & gluon distribution in the photon \\
$\g\g\to J/\psi J/\psi $ & QCD Pomeron 

\end{tabular}
\end{center}
}
\label{processes}
\end{table}
\vspace{-7mm}

\section{Possible luminosities of \GG,\GE\ collisions at TESLA}

As it is well known in \EPEM\ collisions the luminosity is restricted
by beamstrahlung and beam instabilities. In \GG\ collisions these effects are
absent, therefore one can use beams with much smaller cross section.
At present TESLA beam parameters the \GG\ luminosity is
determined only by the attainable geometric \LEE\ luminosity. 

Recently it was found that
horizontal emittance at TESLA damping ring can be reduced by a factor
of 4 in comparison with the previous design.
The resulting parameters of the photon collider at TESLA for 2E=500
GeV and H(130) are presented in Table 2~\cite{GGTEL}. 

\begin{table}[!hbtp]
\caption{Parameters of  the \GG\ collider based on TESLA. 
Left column for 2E=500 GeV, next two columns for Higgs with M=130 GeV, 
two options: $x=1.8, \lambda=1.06$ \MKM\ and $x=4.6 \lambda=0.32$ \MKM.}
{\renewcommand{\arraystretch}{1.2}
\begin{center}
\begin{tabular}{l c c c} 
 & 2E=500 & 2E=200 & 2E=158 \\
 & $x=4.6$ & $x=1.8$ & $x= 4.6$ \\  \hline 
$N/10^{10}$& 2 & 2 & 2  \\  
$\sigma_{z}$, mm& 0.3 & 0.3 & 0.3  \\  
$f_{rep}\times n_b$, kHz& 14.1 & 14.1 & 14.1  \\
$\gamma \epsilon_{x/y}/10^{-6}$,m$\cdot$rad & 2.5/0.03 & 2.5/0.03 & 
2.5/0.03 \\
$\beta_{x/y}$,mm at IP& 1.5/0.3 & 1.5/0.3 & 1.5/0.3 \\
$\sigma_{x/y}$,nm & 88/4.3 & 140/6.8 & 160/7.6  \\  
\LEE(geom), $10^{33}$& 120 & 48 &  38 \\  
$\LGG (z>0.8z_{m,\GG\ }),10^{33} $ & 11.5 & 3.5 &  3.6  \\
$\LGE (z>0.8z_{m,\GE\ }),10^{33}$ & 9.7 & 3.1 & 2.7 \\
\end{tabular}
\end{center}
}
\vspace{-5mm}
\label{table1}
\end{table}   
Figures for the luminosity distribution in \GG\ and \GE\ collisions
can be found elsewhere~\cite{GGTEL}.

For these luminosities the rate of production of the SM Higgs boson
with M$_H$=130(160) GeV in \GG\ collisions is 0.9(3) of that in \EPEM\ 
collisions at 2E = 500 GeV (both reactions, ZH and H$\nu\nu$)~\cite{GGTEL}.

Comparing the \GG\ luminosity with the \EPEM\ luminosity
($L_{\EPEM} = 3\times 10^{34}$ \CMS\ for $2E=500$ GeV)
we see that for the same energy $\LGG(z>0.8z_m) \sim 0.4 L_{\EPEM}.$ 
Having beams with smaller emittances one can get higher \GG\
luminosity, while \EPEM\ luminosity is restricted  by beam
collision effects.

Typical cross sections of interesting processes in \GG\ collisions are
higher than those in \EPEM collisions by about one order of
magnitude~\cite{Tfrei,GGTEL,summary}, so the number of events in \GG\ 
collisions will be more than that in \EPEM\ collisions.  For example,
the cross section for production of $H^+H^-$ pairs in collisions of
polarized photons is higher than that in \EPEM\ collisions by a factor
of 20 (not far from the threshold); this means 8 times higher
production rate for the luminosities given above.

\section{Lasers, optics}

\begin{figure}[!htb]
\centering
\vspace*{0.cm} 
\hspace*{-0.4cm} \epsfig{file=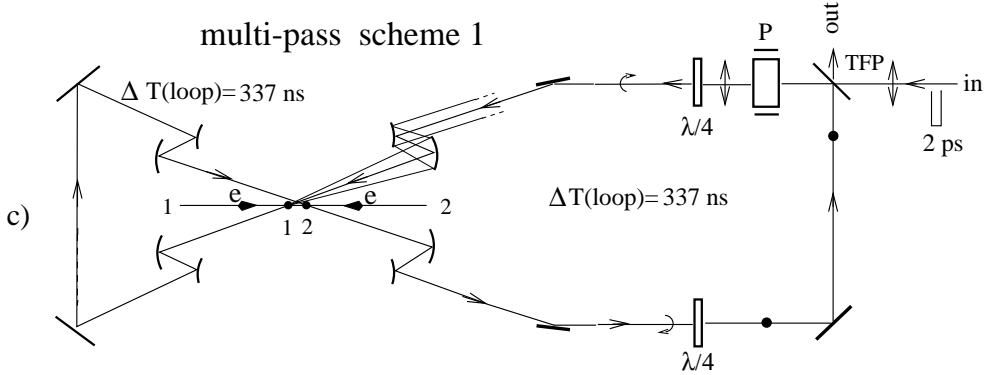,width=13cm,angle=0} 
\vspace*{0.2cm} 
\caption{Optical storage ring for $e \to \gamma$ conversions. 
P is a Pockels cell, TFP is a thin film polarizer, thick dots and 
double arrows show the direction of polarization.}
\vspace{0mm}
\label{loop}
\vspace{-0mm}
\end{figure}

A key element of photon colliders is a powerful laser system which is
used for e$ \to\gamma$ conversion.  Lasers with the required flash
energies (several J) and pulse duration $\sim$ 1 ps already exist.
 The main problem is the high
repetition rate, about 10--15 kHz, with a pulse structure repeating
the time structure of the electron bunches.  
The most attractive and reliable solution at this moment is
an ``optical storage ring'' (fig.\ref{loop}), with a diode pump laser injector.
 This approach can be considered as
a base-line solution for the TESLA photon collider~\cite{GGTEL}.

The laser pulses are send to the interaction region where they are
trapped in an optical storage ring.  Each bunch makes about 6 round
trips (12 collisions with the electron beams) and then is deleted from
the ring. All these tricks can be done by switching one Pockels cell (2 round
trips are possible without Pockels cell).

A laser system required for a such optical storage ring can consist of
about 8 lasers of 1.5 kW average power each.  Due to the high average
power and reliability the lasers should be based on diode pumping. This
technology is developed very actively for a inertial fusion.  Present
cost of diodes for such laser system is about 25 M\$ and it is
expected that their cost will be further decreased several times.
Such system can be done now: all technologies exist.

\section{Conclusion}

The luminosity in \GG\ collisions (in the high
energy peak) can reach about 40\% of \EPEM\ luminosity. Since cross
sections in \GG\ collisions are typically higher by one order of
magnitude than those in \EPEM\ collisions and because of 
access to higher masses for some particles, the photon
collider now has very serious physics motivation.
There is good scheme for the laser system, which, it seems, can be
build now.

\end{document}